\documentclass[preprintnumbers,twocolumn,prl,amsmath,amssymb]{revtex4}

\usepackage{graphicx}
\usepackage{dcolumn}
\usepackage{bm}
\usepackage{SIunits}
\usepackage[french]{babel}

\graphicspath{{Figures/}}

\begin{document}

\title{Magnetic imaging with an ensemble of Nitrogen Vacancy centers in diamond.}
\author{M. Chipaux$^{1}$}
\author{A. Tallaire$^{2}$}
\author{J. Achard$^{2}$}
\author{S. Pezzagna$^{3}$}
\author{J. Meijer$^{3}$}
\author{V. Jacques$^{4}$}
\author{J.-F. Roch$^{4}$}
\author{T. Debuisschert$^{1}$}
\email{thierry.debuisschert@thalesgroup.com}
\affiliation{$^{1}$Thales Research and Technology, 1 av. Augustin Fresnel, F-91767 Palaiseau CEDEX, France}
\affiliation{$^{2}$ Laboratoire des Sciences des Proc\'ed\'es et des Mat\'eriaux, CNRS and Universit\'e Paris 13, 93340 Villetaneuse, France}
\affiliation{$^{3}$ Institut f\"ur Experimentelle Physik II, University Leipzig, Leipzig, Germany}
\affiliation{$^{4}$ Laboratoire Aim\'{e} Cotton, CNRS, Universit\'{e} Paris-Sud and Ecole Normale Sup\'erieure de Cachan, 91405 Orsay, France}

\begin{abstract}
The nitrogen-vacancy (NV) color center in diamond is an atom-like system in the solid-state which specific spin properties can be efficiently used as a sensitive magnetic sensor. An external magnetic field induces Zeeman shifts of the NV center levels which can be measured using Optically Detected Magnetic Resonance (ODMR).
In this work, we quantitatively map the vectorial structure of the magnetic field produced by a sample close to the surface of a CVD diamond hosting a thin layer of NV centers.
The magnetic field reconstruction is based on a maximum-likelihood technique which exploits the response of the four intrinsic orientations of the NV center inside the diamond lattice. The sensitivity associated to a $\unit{1}{\micro\meter\squared}$ area of the doped layer,  equivalent to a sensor consisting of approximately $10^4$ NV centers, is of the order of $\unit{2}{\micro\tesla\per\sqrt{\hertz}}$. The spatial resolution of the imaging device is
$\unit{480}{\nano\meter}$
, limited by the numerical aperture of the optical microscope which is used to collect the photoluminescence of the NV layer. The effectiveness of the method is illustrated by the accurate reconstruction of the magnetic field created by a DC current inside a copper wire deposited on the diamond sample.
\end{abstract}

\maketitle
\section{Introduction}
\label{sec:intro}

\begin{figure*}
\centering
\includegraphics[trim=1.8cm 13cm 1.4cm 1cm, clip=true]{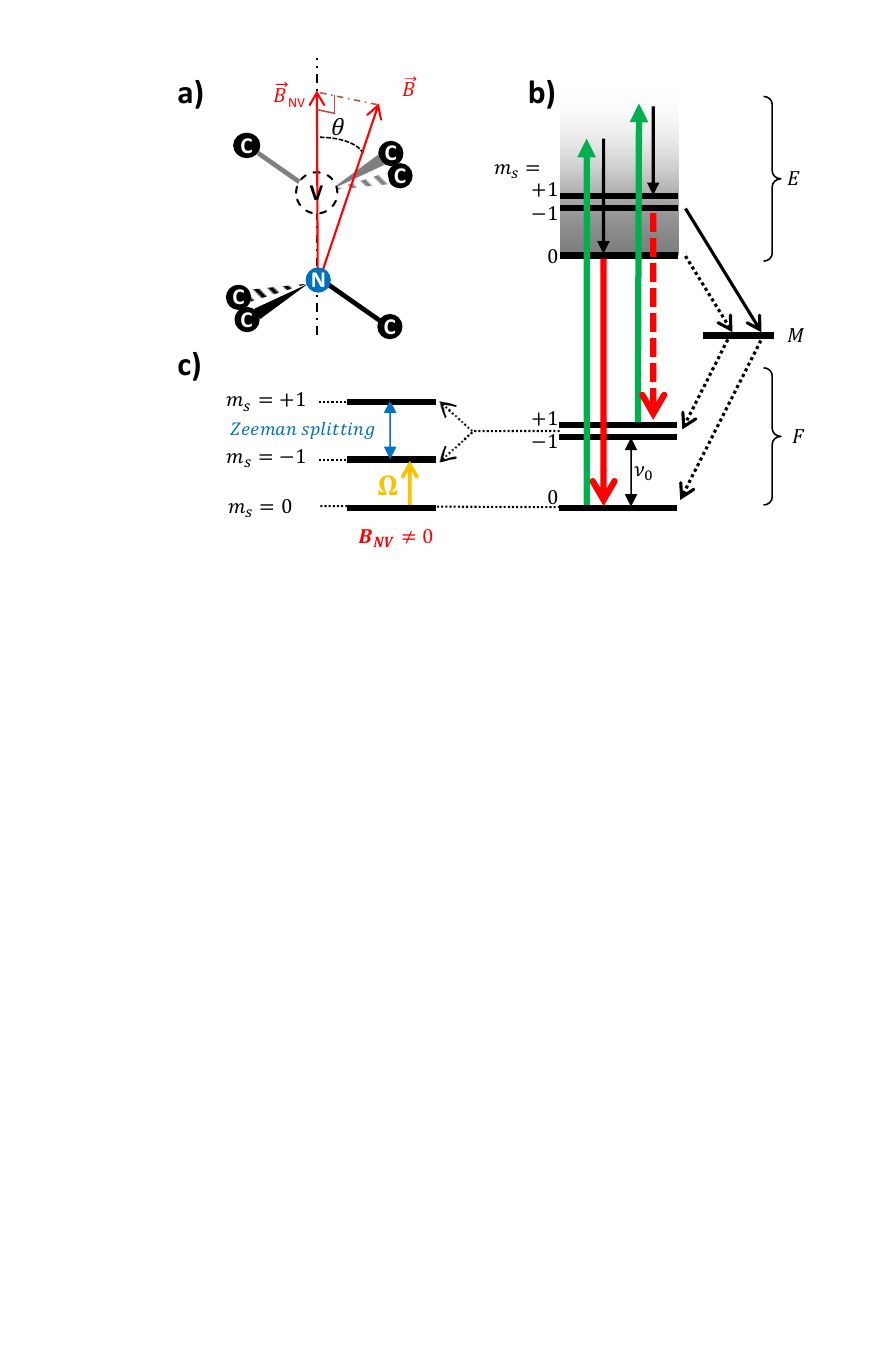}
\caption{\textbf{a)} Structure of the NV center. A Nitrogen atom (N), substituted to a Carbon (C) in the diamond lattice, is coupled to a Vacancy (V) located on an adjacent crystallographic site. The NV center is sensitive to the projection $\vec{B}_{NV}$ of the magnetic field $\vec{B}$ on the N-V axis. The NV$^-$ color centre is a two electrons system which electronic diagram is depicted in \textbf{b)}. It is formed by two spin triplet states denoted F (Fundamental) and E (Excited) and one singlet metastable state M. In the absence of magnetic field, spin states $m_S=+1$ and $m_S=-1$ are lifted from state $m_S=0$ due to spin-spin interaction. The corresponding frequency transition is $\nu_0=\unit{2.88}{\giga\hertz}$ in the fundamental spin triplet. Spin conserving optical transitions can occur between the levels and are represented in green (absorption) and red (emission). Non radiative transitions are represented in black. The $m_S=+1$ and $m_S=-1$ levels are more strongly coupled to the metastable level (solid arrow) than the $m_S=0$ level is (dashed arrow). This dissymmetry explains, first, the possibility to polarize the NV$^-$ center in the $m_S=0$ spin state by optical pumping in the visible (e.g. $\lambda_p=\unit{532}{\nano\meter}$), and, second, the higher rate of photoluminescence when the NV$^-$ center is in spin state $m_S=0$ than when it is in states $m_S=+1$ and $m_S=-1$; \textbf{c)}  In the presence of a magnetic field $\vec{B}$, the degeneracy between state $m_S=+1$ and $m_S=-1$ is lifted proportionally to the field projection on the N-V axis. Those magnetic resonances can be detected applying a microwave field that induces transitions between the states $m_S=0$ and the states $m_S=+1$ or $m_S=-1$ when its frequency $\Omega$ is resonant with the transition frequency. 
The consecutive decrease of the photoluminescence gives rise to optically detected magnetic resonance (ODMR).}
\label{fig:NVCent}
\end{figure*}

Measuring a magnetic field is a generic tool to investigate physical effects involving charge or spin displacement that appear in various fields such as spintronics, nanoelectronics, life-science \cite{Budker2007}. Examples are spin currents in graphene or carbone nanotubes, current propagation in nanoelectronics circuits, neuronal activity inducing a displacement of the action potential.
In addition to sensitivity and spatial resolution, measuring not only the field intensity but also the full vectorial components is particularly valuable, as well as the ability to produce a full image of the sample.

During the past few decades a wealth of methods has been developed to sense and image magnetic fields. Various detection techniques have been investigated such as superconducting quantum interference devices (SQUID) \cite{McDermott25052004}, magnetic resonance force microscopy (MRFM) \cite{Rugar2004,Degen03022009}, alkali vapour atomic magnetometers \cite{Ledbetter19022008,Xu22082006}, and Bose-Einstein condensates \cite{PhysRevLett.98.200801}.

A particularly attractive technique is to develop magnetometers based on Nitrogen Vacancy centers (NV) in ultrapure diamond \cite{Rondin2014_0034-4885-77-5-056503,Schirhagl2014_doi:10.1146/annurev-physchem-040513-103659}. The NV center is a point defect consisting of a substitutional nitrogen atom (N) associated with a vacancy (V) located in an adjacent site of the diamond lattice ({Fig.~\ref{fig:NVCent}~a).
It is as a perfectly photostable 
color center that emits a red fluorescence signal ($\unit{600-800}{\nano\meter}$) when pumped with green light ($\unit{532}{\nano\meter}$).
It is an atom-like system with well defined spin properties.
It can be optically polarized and the fluorescence signal used to detect the spin transition induced by a microwave radiation, thus leading to Optically Detected Magnetic Resonance (ODMR) \cite{Gruber1997}.

NV centers can be produced as single defects with well controlled position within the diamond crystal through ion implantation \cite{Pezzagna2011}. In particular, the depth of the NV centers with respect to the surface can be controlled with a precision of a few nanometers. 
Thus NV centers can be used to realize solid-state, room temperature, optically addressed magnetic sensors \cite{Taylor2008}.

Two main kinds of NV based magnetometers have been developed up to now. Scanning probe magnetometers make use of a monolithic all-diamond scanning probe tip containing a single NV centre within $\sim\unit{10}{\nano\meter}$ from its end \cite{Maletinsky2012}. Sensitivity of $\sim \unit{300}{\nano\tesla\per\sqrt{\hertz}}$
and spatial resolution of $\sim\unit{10}{\nano\meter}$ can be obtained \cite{Grinolds2013}. An alternative solution consists in fixing a nanodiamond containing a single NV at the end of an Atomic Force Microscope tip \cite{Rondin2012}. Such technique has been used recently to visualize domain walls displacement between ferroelectric domains \cite{Tetienne20062014}.

The alternative solution is to use a high-density
ensemble of $N$ NV centres, which should result in a signal to noise enhancement in $\sqrt{N}$ with respect to a single center.
Several devices exploiting that property have been demonstrated \cite{Steinert2010,Pham2011,LeSage2013}. They make use of an active layer of NV centers located close to the surface of a bulk diamond plate. The magnetic object is located close to this surface, and the magnetic field modifies the luminescence emitted by the NV centers. A microscope objective forms a diffraction limited image of the luminescence on a camera. This gives rise to a complete data acquisition over the whole object in one shot.

The goal of the present paper is to describe such a magnetic imaging set-up as well as the reconstruction method of the full vectorial magnetic field.
We first recall the main properties of Nitrogen-Vacancy centers, and in particular those which are specific to ensembles. In a second part, we describe our experimental set-up, which, in particular, involves the total internal reflexion of the pump beam on the faces of the diamond sample.
In a third part we describe how the projections of the field on the crystallographic axes can be obtained from the fluorescence images.
In a fourth part, we detail the maximum-likelihood method that allows for the reconstruction of the magnetic field in each point of the measured image. Finally we study the sensitivity of our set-up and show how it can be optimized using differential acquisition.

\section{Magnetometry with ensemble of NV Centers}

NV centers can exhibit different charge states. The negatively charged NV center (NV$^-$) is a two electrons system with well defined energy levels associated with the spin state (Fig.~\ref{fig:NVCent}~b).
An essential feature of the NV$^-$ defect is that its ground level is a spin triplet $S=1$, which degeneracy is lifted into a singlet state $m_S = 0$ and a doublet state $m_S = \pm1 $, separated by $\nu_0=\unit{2.88}{\giga\hertz}$ in the absence of magnetic field (zero-field splitting) \cite{Manson2006}. The single NV$^-$ center can be polarized in the $m_S = 0$ state by optical pumping and the spin state detected by Optically Detected Magnetic Resonance (ODMR) (see Fig.~\ref{fig:NVCent}~b).

When an external magnetic field is applied, the levels corresponding to $m_S = +1$ and $m_S = -1$ are shifted due to Zeeman effect and their resonance frequencies
denoted $\nu^+$ et $\nu^-$ are shifted accordingly (Fig.~\ref{fig:NVCent}~c). For magnetic fields lower than a few tens 
of Gauss \cite{Tetienne2012}, the Zeeman shift is linear and the positions of the lines are given by:
\begin{equation}\label{eq:Zee1}
	\nu^{\pm}-\nu_0 = \pm \frac{g\mu_b}{h}\cdot B_{NV}\\
\end{equation}
where 
$\nu_0$ is the zero field splitting of the ground state, $\frac{g\mu_b}{h} = \unit{28}{\mega\hertz\per\milli\tesla}$ is the electron spin gyromagnetic ratio and $B_{NV}$ is the projection of the magnetic field along the (N-V) axis (Fig.~\ref{fig:NVCent}~a).
The frequency difference between the two lines $D = \nu^+-\nu^-$ scales linearly with
$B_{NV}$, with the factor:
\begin{equation}\label{eq:Zee2}
	\dfrac{D}{B_{NV}}\simeq \unit{56}{\mega\hertz\per\milli\tesla}
\end{equation}
The NV center has $C_{3v}$ symmetry.
Its symmetry axis can take one of the four crystallographic directions of diamond, which results in four different projections of the magnetic field. 
Therefore, the OMDR spectrum of an ensemble of NV$^-$ centers exhibits four pairs of lines \cite{Steinert2010,Pham2011}. This feature is specific to NV ensembles as compared to single NV and, as detailed in the following, it can be exploited for vectorial reconstruction of the applied magnetic field.

Another point specific to ensembles concerns the sensitivity. When the number of collected photons $N_{phot}$ is high, the signal is shot-noise limited and affected by Poisson noise. Thus the signal to noise ratio varies as $\sqrt{N_{phot}}$ and can be improved increasing the number of collected photons.
This can be done two ways: first, increasing the number of ODMR spectra that are acquired sequentially and second, increasing the number of pixels and thus the corresponding integration area on the
NV center active layer. The number of photons is thus proportional to the product of the integration area $A$ and the integration time $T$.
Therefore, the minimum detectable magnetic field is of the form:
\begin{equation}\label{dBmin}
    \delta B_{min}=\frac{\eta}{\sqrt{T A}}
\end{equation}
where $\eta$ is the sensitivity of the set-up that can be derived measuring $\delta B_{min}$ for various values of $T$ and $A$. In our system $\eta$ is referred to an integration time of $\unit{1}{\second}$ and an integration area of $\unit{1}{\micro\meter\squared}$, and is thus given in ${\micro\tesla\cdot\micro\meter\per\sqrt{\hertz}}$.

\section{Experimental set-up}
\label{sec:Exp}
\begin{figure*}
\centering
\includegraphics[trim=0.8cm 17.6cm 0.6cm 0.2cm, clip=true]{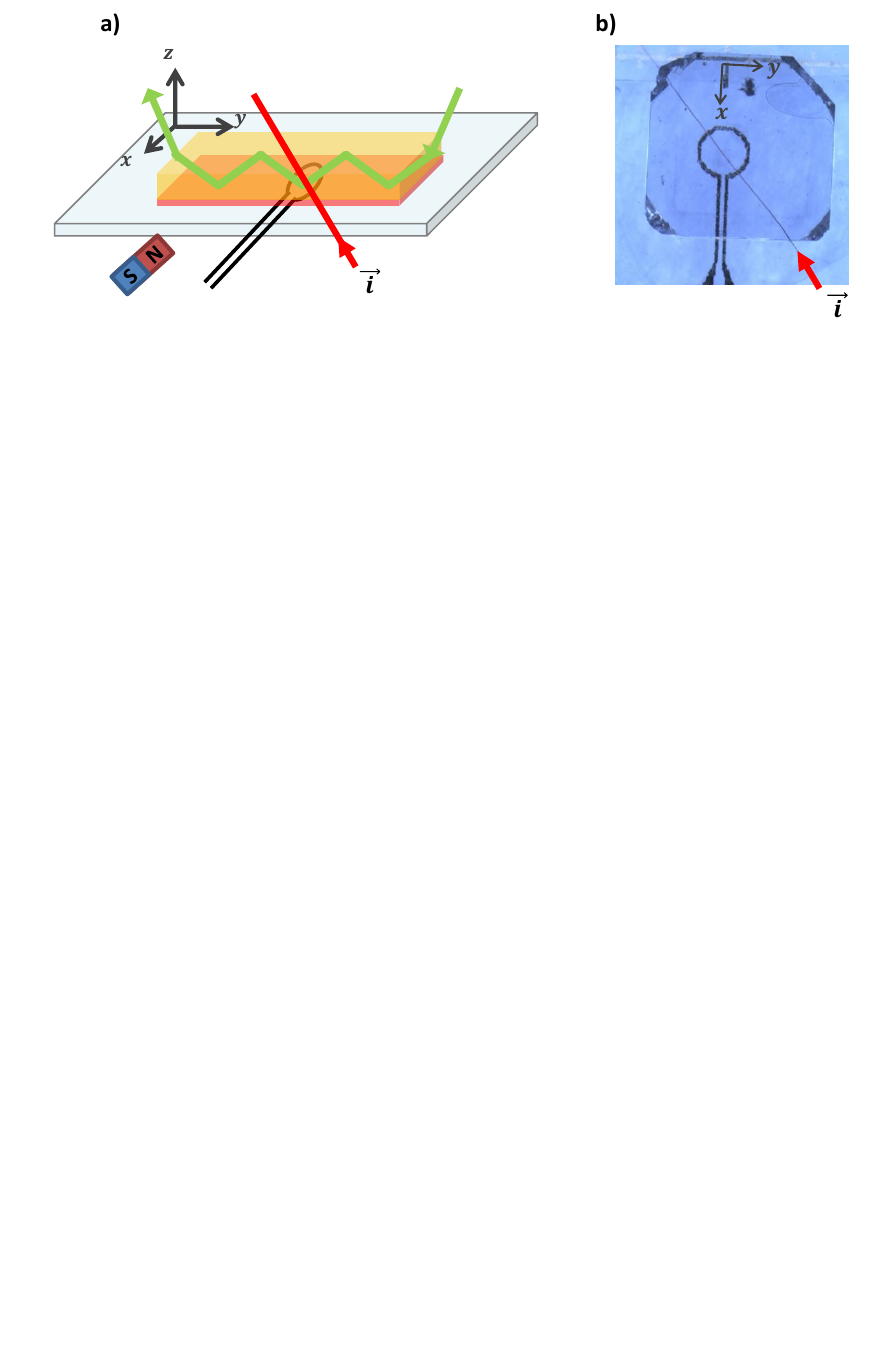}
\caption{\textbf{a)} Scheme of the experiment. The CVD diamond plate (yellow) is implanted with a thin layer of nitrogen vacancy centers close to the surface (transparent red). The pump beam is incident on an optically polished side of the plate and it experiences total internal reflections on the main faces until it reaches the area located close to the sample. The luminescence is emitted by the NV centers in the direction $\left(z\right)$. For the sake of clarity, we have not represented the microscope objective that collects the luminescence towards the CMOS camera. The diamond plate is positioned above a glass plate holding a lithographed short-circuit omega shape antenna allowing to apply the microwave excitation. The $\unit{20}{\micro\meter}$ diameter copper wire holding the DC current is placed between the glass plate and the diamond plate, in contact with the implanted surface. A permanent magnet applies a static magnetic field allowing to lift the degeneracy between the four possible NV centers orientations; \textbf{b)} Picture showing the diamond plate with the antenna and the wire holding the current.}
\label{fig:Settup}
\end{figure*}

The central element of the sensor is the diamond plate holding a thin layer of NV centers implanted close to the surface, represented on {Fig.~\ref{fig:Settup}}.
It is a $(100)$ oriented plate whose edges coincide with axes [100], [010] and [001] and define the laboratory frame denoted (x,y,z) (see {Fig.~\ref{fig:Settup}}). It is obtained from a $(100)$ oriented diamond sample grown by plasma assisted Chemical Vapour Deposition (CVD) \cite{Tallaire2006} on an High Pressure High Temperature (HPHT) substrate. The ultra-pure, single crystal, $\unit{600}{\micro\meter}$ thick sample is then cut to produce a $\unit{3}\times\;\unit{3}{\milli\meter^2}$, $\unit{250}{\micro\meter}$ thick plate. This thickness is chosen to allow the collection of the luminescence of the NV layer located on the back face through the diamond plate, taking into account the $\unit{320}{\micro\meter}$ working distance of the microscope objective. 
This makes possible the study of opaque magnetic samples positioned underneath the plate ({Fig.~\ref{fig:Settup}}). Moreover, this thickness ensures good mechanical stability thus allowing optical quality polishing of the two main faces, which results
in good imaging quality \cite{Almax}.
In addition, the four lateral faces are also optically polished to allow side pumping as detailed in the following.

In order to produce the suitable NV$^-$ layer, the diamond is uniformly implanted \cite{Pezzagna2010} with $^{15}N^{+}$ ions at the density of $\unit{10^{14}}{N\per\centi\meter\squared}$, at the limit of diamond graphitization \cite{Meijer2005_:/content/aip/journal/apl/87/26/10.1063/1.2103389,Prawer200093}. This concentration allows a high density of NV centres while avoiding luminescence quenching by neighboring Nitrogen atoms. The implantation energy is $\unit{5}{\kilo\electronvolt}$, which results in a layer located at about $\unit{8\pm2}{\nano\meter}$ below the surface.
The sample is then annealed at 800~$\degree$C under vacuum for 2 hours to induce vacancy diffusion leading to the conversion of the implanted nitrogen atoms into luminescent NV color centers.
With this energy, around $\unit{1}{\%}$ of the $^{15}N^{+}$ ions are converted in NV$^-$ centres \cite{Pezzagna2010}, 
which results in a surface concentration around $\unit{10^{4}}{NV\per\micro\meter\squared}$. 
Several characterizations performed on different tests samples have shown yield values around $\unit{1}{\%}$ for similar implantation conditions \cite{Pezzagna2010}.

In our set-up, the pump laser is a Coherent~-~Verdi~V5 producing a $\unit{150}{\milli\watt}$ power beam at $\unit{532}{\nano\meter}$.
Two configurations have been considered in order to optically pump the NV$^-$ layer.
The first one, not represented here, consists in focusing the pump beam in the image principal plane of the microscope objective \cite{Steinert2010,Pham2011} in order to obtain a collimated beam on the object side. However, after having pumped the NV$^-$ layer, the beam can heat the sample and may damage it depending on its power.
We have implemented an alternative solution \cite{LeSage2013}, represented in {Fig.~\ref{fig:Settup}~a}, that consists in propagating the pump beam into the diamond sample thanks to total internal reflections on the main faces, taking benefit of the high index of diamond $n_{dia}=2.4$. This technique is made possible thanks to optical quality polishing of the diamond plate side faces.
The pump beam is incident on one side face of the diamond plate, and then experiences zigzag propagation, through total internal reflection, until it reaches the active layer located in the field of view of the microscope objective. Doing this, fragile samples can be studied even with a high power pump beam.

The photoluminescence from the NV$^-$ layer is collected with an immersion microscope objective having a high numerical aperture (N.A = 1.35). This results in a $\unit{8.7}{\%}$ collection efficiency and a diffraction limited resolution of $\unit{480}{\nano\meter}$.
The magnification of the microscope can be chosen by proper selection of its focusing lens. The NV$^-$ layer is imaged on the focal plane of an IDS camera with E2V CMOS sensor. The main characteristics of the imaging system such as
pixel size of the camera, typical magnification value and the corresponding pixel sizes on the sample are given in {Table~\ref{tab:NA}}.

For a typical exposure time of $T = \unit{2.5}{\milli\second}$ a pixel returns a signal close to saturation associated with a signal to noise ratio $\frac{S}{N}$ close to $100$.
Those measured values are consistent with the ones estimated from the sensitivity of the camera.
In addition, we have measured that the signal to noise evolves as $\sqrt{S}$,
which confirms that our measurement is shot-noise limited.

Finally, a ring shape antenna formed by a short-circuit at the end of a coaxial cable powered by a frequency tunable microwave synthesizer provides an oscillating microwave magnetic field that is used to induce the magnetic resonances of the NV centres.

\begin{table}
\caption{Main characteristics of the pump beam, the imaging system and the measured OMDR lines.}
\label{tab:NA}
\begin{tabular}{lll}
\hline\noalign{\smallskip}
Parameter 						& symbol 			& value\\
\noalign{\smallskip}\hline\noalign{\smallskip}
Pumping wavelength 				& $\lambda_{pump}$ 	& $\unit{532}{\nano\meter} $\\
Pumping power					& $P_{pump}$		& $\unit{150}{\milli\watt}$\\
\noalign{\smallskip}\hline\noalign{\smallskip}
Exposure time 					& $T$ 				& $\unit{2.5}{\milli\second}$ \\
Size of the CMOS array pixels 	& $e'$				& $\unit{5.3}{\micro\meter}$\\
Microscope magnification		& $G$				& $\unit{25}{}$\\
Pixel size on the active layer	& $e'/G$			& $\unit{210}{\nano\meter}$\\
\noalign{\smallskip}\hline\noalign{\smallskip}
Signal to noise ratio (one pixel)& $S/N$ 			& $\unit{105}{}$\\
Contrast of ODMR lines			& $C$				& $\unit{0.9}{\%}$\\
Linewidth (at half maximum) 	& $\Delta\nu$		& $\unit{6.8}{\mega\hertz}$\\
\noalign{\smallskip}\hline\noalign{\smallskip}
\end{tabular}
\end{table}

\section{Data acquisition and treatment}
\label{sec:acq}
\begin{figure*}
\centering
\includegraphics[trim=0.5cm 4.7cm 0.4cm 0.5cm, clip=true]{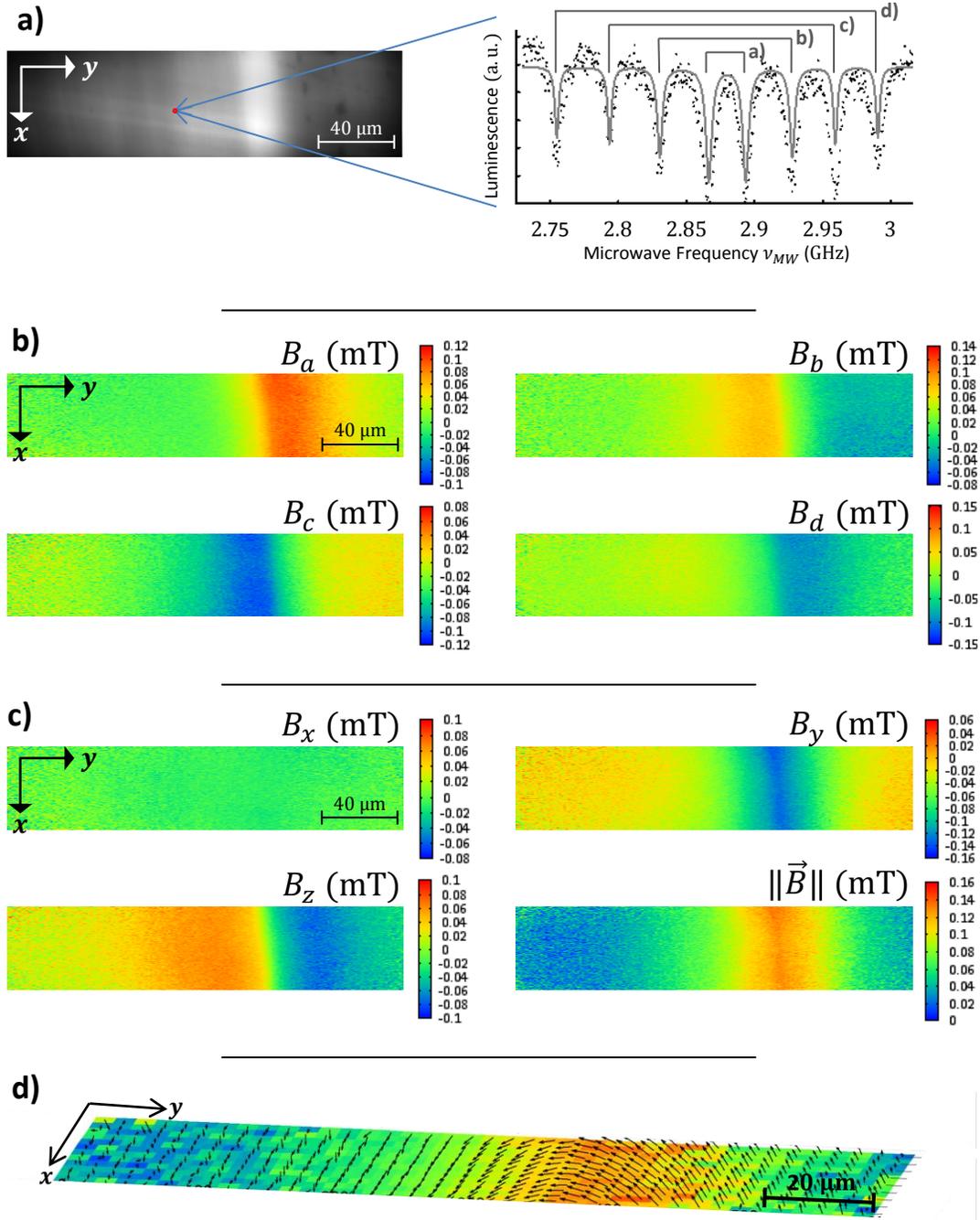}
\caption{Illustration of the consecutive steps leading to the vectorial reconstruction a magnetic field with our set-up. The magnetic field is produced by a current of $\unit{12}{\milli\ampere}$ within a copper wire located underneath the diamond plate (Fig.~\ref{fig:Settup}).
The field is measured in the plane of the NV$^-$ layer.
The frequency span is equal to $\unit{300}{\mega\hertz}$ with a sampling of $\unit{0.5}{\mega\hertz}$. The signal is averaged over $64$ sweeps
; \textbf{a)} Image of the luminescence from the NV$^-$ centres layer obtained at one given frequency and OMDR spectrum measured for a pixel located at the centre of the image. The raw data are composed of a full ODMR spectrum for each pixel;
\textbf{b)} Projection of the magnetic field
along the four crystallographic directions $\left(a\right)$, $\left(b\right)$, $\left(c\right)$ and $\left(d\right)$ for each point in the plane calculated from the raw data by the fitting algorithm;
\textbf{c)} Reconstruction of the magnetic field along each axis $\left(x,y,z\right)$ of the laboratory frame
({cf.~Fig.~\ref{fig:Settup}}) using the maximum likelihood method and norm of the magnetic field;
\textbf{d)} Vectorial representation of the magnetic field produced by the sample in the laboratory frame.}
\label{fig:EsrImg}
\end{figure*}

The set-up described above allows to perform an image of the NV$^-$ layer, and for each pixel, to retrieve the ODMR spectrum. The frequency of the microwave synthesizer is swept around the central frequency $\nu_0=\unit{2.88}{\giga\hertz}$.
For each frequency step, a complete luminescence image is taken (Fig.~\ref{fig:EsrImg}~a). The images are appended in the computer memory to form a 3D volume of data giving the value of the luminescence for each pixel (x,y) of the camera and for each frequency ($\nu$) of the sweep. This volume of data can also be considered as an image in which each (x,y) pixel returns a full ODMR spectrum. {Fig.~\ref{fig:EsrImg}~a} presents such a spectrum taken from a pixel at the centre of the image.
A static magnetic field is applied to shift the lines away from degeneracy. The spectrum exhibits four pairs of lines, denoted $(a)$, $(b)$, $(c)$ and $(d)$ corresponding to the four possible orientations of the NV$^-$ centre.
Each pair $(i)$ corresponds to the frequencies $\nu_i^+$ and $\nu_i^-$ located symmetrically each side of the central frequency $\nu_0$. The distance between the two peaks in one pair is directly related to the projection $B_i$ of the applied magnetic field on the corresponding N-V axis (Eq.~\ref{eq:Zee2}).

The precise position of the eight resonances are determined for each pixel using the Levenberg-Marquardt algorithm \cite{Levenberg1944,Marquardt1963} to fit the following multi-Lorentzian, multi-parameter function on the data:

\begin{multline}\label{eq:ffit}
	f(\nu) = y_0\times
	\left(
		1 - \sum_{\substack{i=\left(a\right)\cdots\left(d\right)}}C_i\cdot
		\left[
			L\left(
				\dfrac{\nu-\bar{\nu_i}-D_i/2}{\Delta\nu_i}
			\right)\right.\right.\\
			\left.\left. + L\left(
				\dfrac{\nu-\bar{\nu_i}+D_i/2}{\Delta\nu_i}
			\right)
		\right]
	\vphantom{\sum_{\substack{i=\left(a\right)\cdots\left(d\right)}}}\right)
\end{multline}

where $L\left(x\right)=\dfrac{1}{1+x^2}$ is the Lorentzian function, $C_i$ is the contrast of the line's pair $(i)$, $\bar{\nu_i}=\dfrac{\nu_i^-+\nu_i^+}{2}$ is the central position of the pair, $D_i=\left|\nu_i^+-\nu_i^-\right|$ is the Zeeman distance between the two lines of the pair and $\Delta\nu_i$ is their linewidth.

The algorithm requires inputs for the fitting parameters that are not too far from the result. So, a pre-selection of those parameters is performed manually for a pixel in the centre of the image. Then the entire image is fitted gradually taking the results of the neighboring already fitted pixels as an input for the following one.

At the end $D_a$, $D_b$, $D_c$ and $D_d$, the distance between the two transition for each class of NV centre, are obtained for each pixel of the camera.
As a convention, $(a)$, $(b)$, $(c)$ and $(d)$ are chosen in the following order:

\begin{equation}\label{eq:ineq}
 0 < D_a < D_b < D_c < D_d
\end{equation}

The measurements of the magnetic field $\vec{B}$ projected on the four possible NV orientations are denoted $m_a$, $m_b$, $m_c$ and $m_d$. Their absolute values are derived using Eq.~(\ref{eq:Zee2}) and Eq.~(\ref{eq:ineq}).
The sign of those projections and the orientation of the axes $(a)$, $(b)$, $(c)$ and $(d)$ with respect to the laboratory frame are then to be determined.

$\vec{u_a}$, $\vec{u_b}$, $\vec{u_c}$ and $\vec{u_d}$ are the unit vectors representing the four possible orientations of the NV axes. Due to the symmetry properties of the NV centres, they are related by:
\begin{equation}
\vec{u_a} + \vec{u_b} + \vec{u_c} + \vec{u_d} = \vec{0}
\end{equation}
Taking the projection of this equation along the field $\vec{B}$ gives:
\begin{equation}\label{eq:sum0}
B_a + B_b + B_c + B_d = 0
\end{equation}

Both field $\vec{B}$ and $-\vec{B}$ have 
the same signature on the NV centres and cannot be distinguished with our system. We choose $B_d > 0$.
Then $B_b$, $B_c$ verifying both {Eq.~(\ref{eq:ineq})} and {Eq.~(\ref{eq:sum0})} have necessarily a sign opposite to that  of $B_d$.
In the absence of noise, $B_a$ should be directly deduced from {Eq.~(\ref{eq:sum0})}. In the presence of noise, we choose the sign $\epsilon$ of $B_a$, minimizing the relation $\epsilon\cdot D_a -D_b - D_c + D_d $. With this method, we can most probably find the good values of the field projections $m_i$ from the measured $D_i$.

In practice, the fitting algorithm is robust. It returns a reliable value if the contrast $C$ exceeds approximatively twice the noise
fluctuations i.e. when the ODMR line is hardly visible from the noise.
Considering a typical contrast of $\unit{1}{\%}$ and a signal to noise ratio for a single sweep close to $100$ {(Table~\ref{tab:NA})} an average of only four sweeps is sufficient to successfully retrieve the experimental parameters. The fitting algorithm works with a minimum of $3$ or $4$ frequency samples within an ODMR line. So considering a typical linewidth of $\Delta\nu=\unit{7}{\mega\hertz}$, a sampling resolution of $\unit{2}{\mega\hertz}$ is sufficient. As a result, for a typical frequency span of $\nu_{span}=\unit{300}{\mega\hertz}$, 150 frequency samples (i.e. images) are necessary. Considering a minimum number of 4 sweeps and an exposure time of 2.5 ms, the minimum duration for the entire measurement is around $\unit{1.5}{\second}$ for a spatial sampling of one pixel on the camera, corresponding to $\unit{210}{\nano\meter}$ on the diamond plate. Instead of accumulating several frequency sweeps, the signal to noise ratio can be increased by spatial binning of the camera pixels. A binning of $\unit{5\times5}{pixels}$, corresponding to a $\unit{1}{\micro\meter}$ spatial resolution on the diamond plate, requires only one frequency sweep and an acquisition time around $\unit{300}{\milli\second}$ to result in a signal to noise ratio equivalent to the one obtained in the previous case. Therefore, an optimization of the spatial resolution / acquisition time can be performed.

\section{Magnetic field reconstruction}
\label{sec:MagRec}

\begin{figure*}
\centering
\includegraphics[trim=0.5cm 15cm 0cm 0.5cm, clip=true]{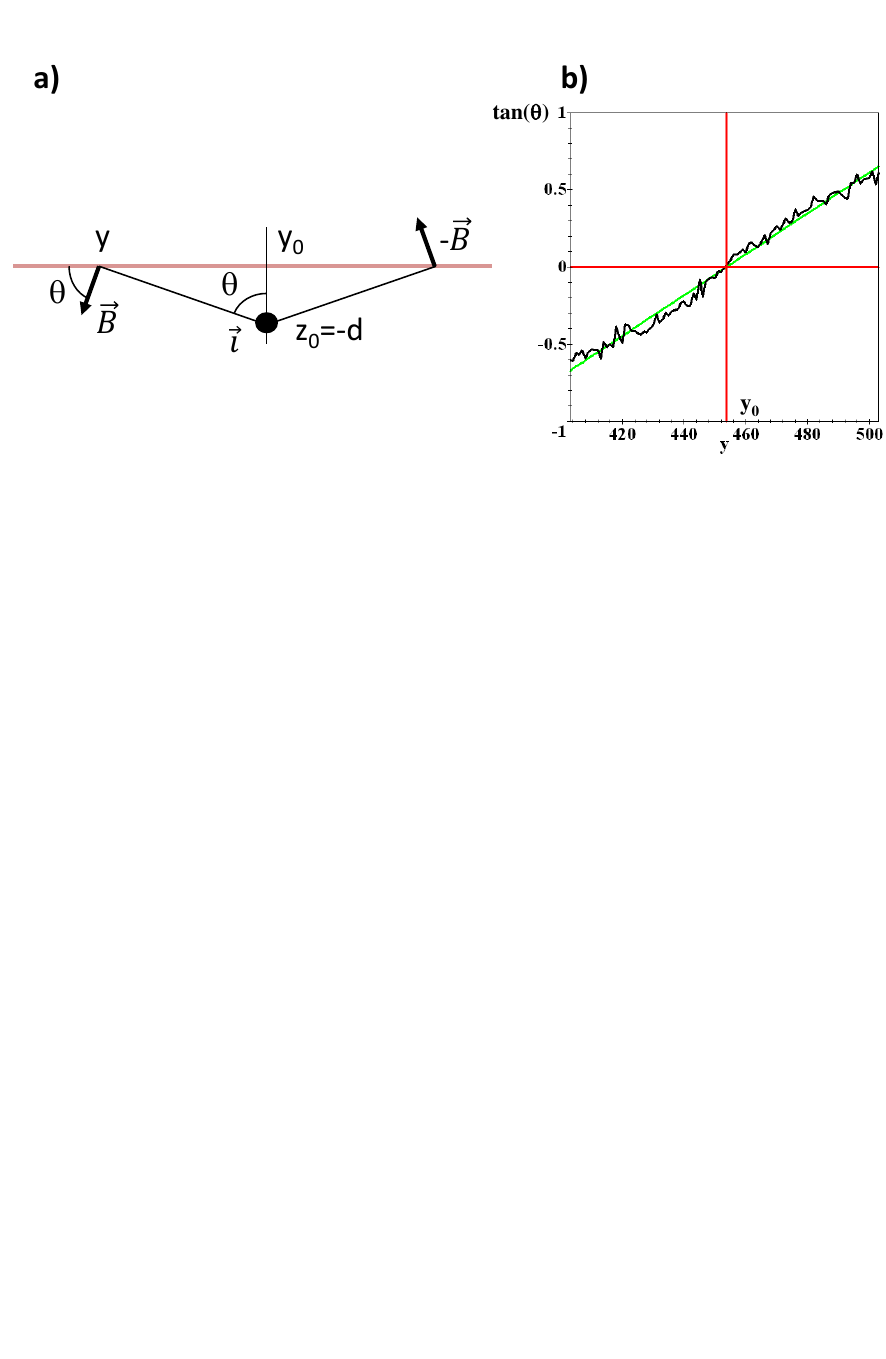}
\caption{\textbf{a):} Scheme of the magnetic field produced on the NV center layer (pink) at location (y,z=0) by a current perpendicular to the plane located in ($y_0,z_0$).
The angle between the magnetic field $\vec{B}$
and the NV layer is denoted $\theta$.
It is equal to zero when $y=y_0$. \textbf{b):} Dependence of $tg(\theta)$ as function of $y$. A linear fit(green) allows to retrieve the position of the wire ($y_0,z_0$).
}
\label{fig:Fil}
\end{figure*}

Once the projections of the magnetic field on the crystallographic axes have been determined, the following step is to reconstruct the magnetic field in the laboratory frame.  The measured field components $m_i$ do not directly give the magnetic field $B_i$, but are affected by some noise inherent to the measurement. To account for that effect, we retrieve the best evaluation of the magnetic field knowing the values of  $m_i$ using a maximum-likelihood method.

We define the frame $\left(x_1, y_1, z_1\right)$ in which $\vec{u_a}$, $\vec{u_b}$, $\vec{u_c}$ and $\vec{u_d}$ have the following coordinates:

\begin{equation}\label{eq:Defa}
\begin{array}{cc}

\vec{u_a} = \frac{1}{\sqrt{3}}
\begin{pmatrix}
	-1 \\ -1 \\ -1
\end{pmatrix}_{\left(x_1, y_1, z_1\right)}
&
\vec{u_b} = \frac{1}{\sqrt{3}}
\begin{pmatrix}
	1 \\ 1 \\ -1
\end{pmatrix}_{\left(x_1, y_1, z_1\right)}
\\
\vec{u_c} = \frac{1}{\sqrt{3}}
\begin{pmatrix}
	-1 \\ 1 \\ 1
\end{pmatrix}_{\left(x_1, y_1, z_1\right)}
&
\vec{u_d} = \frac{1}{\sqrt{3}}
\begin{pmatrix}
	1 \\ -1 \\ 1
\end{pmatrix}_{\left(x_1, y_1, z_1\right)}
 \\
\end{array}	
\end{equation}

Due to the (100) orientation of the diamond, each axis $\left(x_1\right)$, $\left(y_1\right)$ and $\left(z_1\right)$ coincides with the edges of the plate ({Fig.~\ref{fig:Settup}~a). However, they are still to be attributed to each axis $\left(x\right)$, $\left(y\right)$ and $\left(z\right)$ of the laboratory frame since the orientations of the crystalline axes $\left(a\right)$, $\left(b\right)$, $\left(c\right)$ and $\left(d\right)$ are chosen with respect to $\overrightarrow{B}$ (cf. Eq. \ref{eq:ineq}), which can have an arbitrary direction.

We consider a Gaussian noise distribution with standard deviation $\sigma$. The probability to measure $m_a$ along axis $(a)$ knowing $\vec{B}$ and thus $B_a$ is given by:

\begin{equation}
p\left(m_a|\vec{B}\right)\propto e^{\frac{-{\left(B_a-m_a\right)}^2}{2\sigma^2}}
\end{equation}
Thus, the probability to measure $m_a$, $m_b$, $m_c$ and $m_d$ knowing $\vec{B}$ is:

\begin{align}\label{proba}
p\left(\{m_i\}|\vec{B}\right) = \prod_{i=\left(a\right)\cdots\left(d\right)} p\left(m_i|\vec{B}\right) & \propto \exp{\left(-E\right)}
\end{align}

with $E$ given by
\begin{equation}
E = \sum_{i=\left(a\right)\cdots\left(d\right)}\left[\frac{1}{2\sigma^2}{\left(B_i-m_i\right)}^2\right]
\end{equation}
According to Bayes theorem, we can interpret Eq.~(\ref{proba}) as the likelihood to have a magnetic field equal to $\vec{B}$, knowing the actual measurements $m_a$, $m_b$, $m_c$ and $m_d$. Thus, the best estimation of $\vec{B}$ would minimize $E$.

Therefore, we express the projections of the magnetic field on the $(i)$ direction, $B_i=\vec{B}\cdot\vec{u_i}$, as functions of $B_{x_1}$, $B_{y_1}$ and $B_{z_1}$ and substitute these expressions in $E$.
Then we minimize $E$ with respect to $B_{x_1}$, $B_{y_1}$ and $B_{z_1}$ 
and obtain the expressions of the most likely magnetic field components as a function of $m_a$, $m_b$, $m_c$ and $m_d$:

\begin{equation}
\begin{array}{c}
B_{x_1} = \frac{\sqrt{3}}{4}\left(-m_a + m_b - m_c + m_d\right)\\
B_{y_1} = \frac{\sqrt{3}}{4}\left(-m_a + m_b + m_c - m_d\right)\\
B_{z_1} = \frac{\sqrt{3}}{4}\left(-m_a - m_b + m_c + m_d\right)
\end{array}
\end{equation}

The advantage of those expressions is to involve the four measured projections of the field, $m_i$, and, thus, to exploit all the available information. They give the best estimate of $\overrightarrow{B}$, even if the measurement is affected by noise.

The last step is to find the proper permutation of axes $\left(x_1\right)$, $\left(y_1\right)$, $\left(z_1\right)$ in order to retrieve the components of the magnetic field in the laboratory frame.
This can be solved exploiting prior knowledge on the sample. Another possibility is to add an auxiliary known magnetic field (for example by adding a CW current in the antenna), that allows identifying the four lines $(a)$, $(b)$, $(c)$ and $(d)$ with respect to the laboratory frame.

As an example, {Fig.~\ref{fig:EsrImg}} displays such a reconstruction for a magnetic field produced by a current of $\unit{12}{\milli\ampere}$ in a $\unit{20}{\micro\meter}$ diameter copper wire. In this case the shape of the magnetic field distribution that is expected is known a priori which allows to attribute the good directions to the magnetic field components. 

Once the magnetic field is known, it can be used to retrieve the characteristics of the source. In the case of a wire, this can be done by simple geometrics considerations.
Up to small misalignments, the wire is aligned with $x$ axis.
The circulating current produces an ortho-radial magnetic field located mainly in the $(y,z)$ plane. The lines perpendicular to the fields in each position $(y,0)$ along the NV center layer are characterized by the angle $\theta$ and intersect at the location of the wire $(y_0,z_0)$.
Knowing $B_z$ and $B_y$, we express $\tan(\theta)$ at each position $(y,0)$:
\begin{equation}\label{theta}
    \tan(\theta)=\frac{B_z}{B_y}=\frac{\left(y-y_0\right)}{d}
\end{equation}
A linear fit of the data intersects the $\tan(\theta)=0$ line  and gives the value of $y_0$. The inverse of the slope gives the value of $d$. As a consequence the wire can be positioned with respect to the NV center layer. Considering the field measurements on one side of the field $(x=0)$ we can extract the values $y_0=454$ and $d=76$ expressed in pixels units. Taking into account the effective size of the pixel on the NV layer (Table \ref{tab:NA}), we obtain $y_0$ = 95.3 $\mu$m and d = 15.9 $\mu$m. This value is consistent with the 20 $\mu$m nominal diameter of the copper wire, also holding an additional isolating cover. The difference between the measured distance and the wire radius can be explained by the fact that the wire is not sticked to the diamond plate.

In addition, measuring the norm of the magnetic field allows to estimate the current circulating in the wire. For a position $(y,0)$ on the NV centers layer, the current is given by
\begin{equation}\label{current}
    i(y)=\mu_0^{-1}B(y) 2 \pi \sqrt{z_0^2+(y_0-y)^2}
\end{equation}
Averaging the values obtained in the vicinity of the magnetic field maximum gives a result of 10.5 mA. Altough a little lower than the expected value (12 mA), this value shows that a good quantitative agreement can be obtained from our simple method.

We have performed the same evaluation on the other side of the field located at x = 37.4 $\mu$m (178 pixels). We obtain a wire position given by $d$ = 16.0 $\mu$m (76 pixels) and $y_0$ = 98.6 $\mu$m (469 pixels). This shows that the wire is almost perfectly parallel to the NV layer and that there is a small tilt angle of 5 $\degree$ with respect to x axis. We have calculated a 10.2 mA current in the wire for that position, close to the one obtained on the other side of the sample. The slight difference is due to the simplifications in our method that considers two independent (y,z) planes. A more precise method should involve one uniform current in the wire all along the sample and exploit all available magnetic field measurements.

This simple reconstruction method shows that our technique is suited for retrieving the characteristics of simple objects such as a wire. It could be extended further to study the characteristics of more complicated source distributions.

\section{Sensitivity}
\label{sec:Sens}

\begin{figure*}
\centering
\includegraphics[trim=0.5cm 16cm 0.5cm 0.5cm, clip=true]{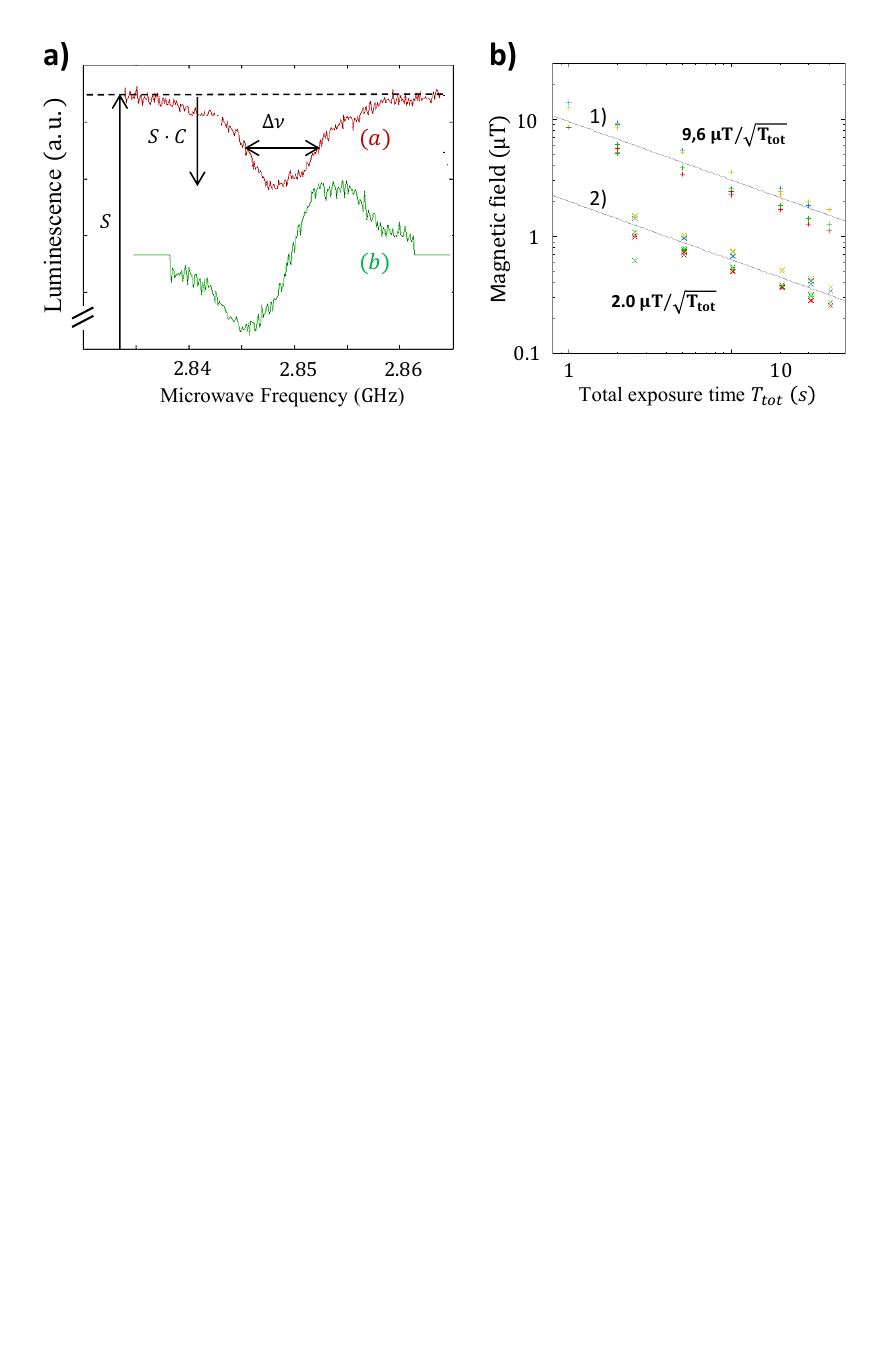}
\caption{\textbf{a)} The red curve (a) represents a typical ODMR line.
The total luminescence signal is $S$. The amplitude of the resonance is $S\times C$ where $C$ is the contrast. The linewidth is $\Delta\nu$.
The green curve (b) is the error function $Err(\nu)$ obtained by a differential acquisition (Eq.~(\ref{Err})) derived from the OMDR curve (a);
\textbf{b)} Minimum detectable magnetic field, for a 1 $\mu m^2$ acquisition area, as a function of the total acquisition time. Two cases are represented. The line $1)$ is obtained with a multi-Lorentzian fitting algorithm. The line $2)$ is obtained with the differential acquisition method. Those lines allow to retrieve the sensitivities of both methods.}
\label{fig:Noise}
\end{figure*}

In the previous part, we have depicted a method allowing a vectorial reconstruction of the magnetic field starting from the measurement of the field projections on the crystallographic axes. Here, we evaluate the sensitivity of this technique and estimate the maximum sensitivity of our set-up according to our experimental parameters.

We first determine experimentally the efficiency of the fitting algorithm. 
Therefore, we measure two times the same uniform magnetic field consecutively. We then subtract the data and perform a statistics over a large number of $32\times32$ pixels. The resulting variance is two times the variance of one measurement $\delta B_{fit}^2$.
The minimum detectable magnetic field $\delta B_{fit}$, normalized to a 1 $\mu m^2$ integration area, is given in {Fig.~\ref{fig:Noise}~b} as a function of the total acquisition time $T_{tot}$. We have limited our acquisition time to 10 s, which is indeed long with respect to many physical phenomena. Over this time-scale, the minimum detectable magnetic field features a shot-noise scaling. On the other hand, we investigated how the signal-to-noise ratio can be improved by increasing the integration surface. Starting from one pixel, we found that the signal-to-noise ratio has a shot-noise dependence until the integration surface reaches 8 x 8 pixels where it starts to saturate. For a pixel number around 1500, increasing further the number of integrated pixels does not bring any advantage anymore. This is probably due to the presence of technical noise in the camera response that induces correlated fluctuations between the pixels. Of course, this increase of signal-to-noise ratio by pixel integration is obtained at the price of spatial resolution.

From line $1)$ in {Fig.~\ref{fig:Noise}~b}, we retrieve the normalized sensitivity of the fitting algorithm:

\begin{equation}\label{eq:etafit}
\begin{array}{rl}
	\eta_{fit}&=\unit{9.6}{\micro\tesla\cdot\micro\meter\per\sqrt{\hertz}}
\end{array}
\end{equation}

Then, we evaluate the maximal sensitivity that can be expected from our set-up. It is determined from an ODMR line such as the curve (a) of {Fig.~\ref{fig:Noise}~a}. The total luminescence signal is $S$. The amplitude of the resonance is $S\times C$ where $C$ is the contrast. The linewidth is $\Delta\nu$.
The maximum of the slope is obtained close to the line half-maximum and is equal to $\dfrac{1}{0.77}\cdot\dfrac{C}{\Delta\nu}$. The factor $0.77$ comes from the  shape of the line that is assumed Lorentzian.
The noise
is the standard deviation of the monitored signal $S$.

The sensitivity is calculated from the parameters values obtained for one pixel of the camera that are given in {Table~\ref{tab:NA}}. We obtain \cite{Taylor2008,Maze2008}:

\begin{equation}\label{eq:bmin}
\begin{array}{rl}
	\eta_{max} 	&= \dfrac{0.77}{g\mu_b/h}\cdot\dfrac{\Delta\nu}{C\frac{S}{N}}\cdot\sqrt{T\cdot A}\\
			&= \unit{2.0}{\micro\tesla\cdot\micro\meter\per\sqrt{\hertz}}
\end{array}
\end{equation}

This value is better than the one obtained with the fitting algorithm given by Eq.~(\ref{eq:etafit}). The ratio between those two values is given by $\sqrt{\dfrac{2\Delta\nu}{\nu_{span}}}$ , which shows that, over a complete spectrum given by $\nu_{span}$, only the part corresponding to the two resonance lines necessary to retrieve the magnetic field, $2\Delta\nu$, actually brings useful information on that field.

In order to optimise the measurement time, all the acquisitions have to be taken in the frequency range where the slope of the line is maximum. Having a prior estimation of its position $\nu$, we can acquire two images at frequencies $\nu_a=\nu+\Delta\nu/2$ and $\nu_b=\nu-\Delta\nu/2$. The difference $S\left(\nu_a\right)-S\left(\nu_b\right)$ is immune from the common mode noise and proportional to the shift of the magnetic field with respect to the central position $\nu$ \cite{Schoenfeld_PhysRevLett.106.030802,Rondin2012}. A normalisation by $S\left(\nu_a\right)+S\left(\nu_b\right)$ cancels the spatial variation of the pumping beam intensity. We can then calculate the error function
\begin{equation}\label{Err}
    Err(\nu)=\dfrac{S(\nu_a)-S(\nu_b)}{S(\nu_a)+S(\nu_b)}
\end{equation}
that is represented by the curve (b) of {Fig.~\ref{fig:Noise}~a}.
Knowing the contrast and the linewidth, we can retrieve the minimum detectable magnetic field from the measured signal as a function of the total acquisition time $T_{tot}$.
It is fitted by the line $2)$ in {Fig.~\ref{fig:Noise}~b}. The resulting sensitivity is
\begin{equation}\label{etadiff}
    \eta_{diff}=\unit{2.0}{\micro\tesla\cdot\micro\meter\per\sqrt{\hertz}}
\end{equation}
which matches exactly the value of $\eta_{max}$. Therefore this differential method allows to obtain the maximum sensitivity of our set-up once the line positions have been determined by the fitting algorithm. Another advantage of this differential method is that the measurement of the slope requires only two frequency measurements and thus two images, leading to
much shorter acquisition times.

The contrast and the resonance linewidth are essential parameters determining the sensitivity. We have measured contrasts in the range of $\unit{1}{\%}$ to $\unit{2}{\%}$. Those values are smaller than those obtained with single NV centers (typ. $\unit{20}{\%}$). This difference can be explained from several origins. 
First, we collect at the same time the luminescence from 4 sets of NV centers corresponding to the 4 possible orientations. Only one set is resonant at a time, the three others result in a background signal that leads to a factor four decrease of the contrast value as compared to single NV centers.
Second, our sample includes neutral NV centers (NV$^0$) as well which contribute to the background luminescence.
Third, the pumping intensity is necessarily much lower with ensemble, due to the fact that a large surface is to be pumped. Thus, the
polarization in the $m_S=0$ state is lower. 
Finally, our sample is heavily doped with N ions. Thus each NV center is surrounded with a high density electron spin bath that can result in a short relaxation time and a decrease of the
polarization rate.
The combination of those various effects can explain the low contrast obtained with our set-up.

Several improvement directions can be foreseen. Preferential NV orientation can be obtained thanks to controlled N doping during the growth of diamond over specific orientations \cite{Lesik2014_:/content/aip/journal/apl/104/11/10.1063/1.4869103,Michl2014_:/content/aip/journal/apl/104/10/10.1063/1.4868128,Fukui2014_1882-0786-7-5-055201}.
In addition, controlling the charge state with suited surface termination \cite{:/content/aip/journal/apl/96/12/10.1063/1.3364135} would allow to increase the rate of useful NV$^-$ centers as compared to NV$^0$ centers.

The observed linewidth is in the range of 7 MHz.
One possible explanation of this linewidth can come from the power broadening due to continuous pumping. Using pulsed measurement protocols \cite{Taylor2008} can avoid that effect and result in sharper lines as has been observed with single NV centers with linewidth in the range of 100 kHz \cite{Dreau2011}.
Line broadening can also come from the electron spin bath resulting from the high quantity of Nitrogen atoms that are not converted into NV centers.
Increasing the yield of the conversion from N to NV using techniques such as electron or proton irradiation would allow to work with a lower concentration of Nitrogen and thus decrease this effect.
Ultimately, $^{12}C$ isotopically enriched diamond can eliminate the nuclear spin bath due to $^{13}C$ that are present in natural diamond.

\section{Conclusion}
\label{Sec:Disc}

In this work, we have exploited the ODMR signal of an ensemble of NV centers in order to quantitatively map the vectorial structure of a magnetic field produced by a sample close to the surface of a CVD diamond hosting a thin layer of NV centers. The reconstruction of the magnetic field is based on a maximum-likelihood technique which exploits the response of the four intrinsic orientations of the NV center inside the diamond lattice. The sensitivity associated to a $\unit{1}{\micro\meter\squared}$ area of the doped layer,  equivalent to a sensor consisting of approximately $10^4$ NV centers, is of the order of $\unit{2}{\micro\tesla\per\sqrt{\hertz}}$. The spatial resolution of the imaging device is $\unit{480}{\nano\meter}$,
limited by the numerical aperture of the optical microscope which is used to collect the photoluminescence of the NV layer.
The effectiveness of this technique is illustrated by the accurate reconstruction of the magnetic field created by a DC current inside a copper wire deposited on the diamond plate. We have investigated the limitations that are either specific to our technique or related to the physical properties of our diamond crystal. We have proposed several improvement directions that should allow to reach sensitivities in the ${\nano\tesla\per\sqrt{\hertz}}$ range for an integration area of $\unit{1}{\micro\meter\squared}$.

\section{Acknowledgements}
\label{Sec:Ack}

The research leading to these results has received funding from the European Union
Seventh Framework Programme ({FP7/2007-2013}) under the project DIADEMS ({grant agreement} n$^{\circ}$ 611143) and from the Agence Nationale de la Recherche (ANR) under the project ADVICE (grant ANR-2011-BS04-021).

\bibliographystyle{unsrt}

\end{document}